\documentclass[12pt,preprint,useAMS,usenatbib]{emulateapj}
\usepackage{graphicx}
\usepackage{subfloat}
\usepackage{amsmath}

\def\msun{\,{M_\odot}}

\def\Rvir{R_{\rm vir}}
\def\Mvir{M_{\rm vir}}

\def\spose#1{\hbox to 0pt{#1\hss}}
\def\lta{\mathrel{\spose{\lower 3pt\hbox{$\mathchar"218$}}
     \raise 2.0pt\hbox{$\mathchar"13C$}}}
\def\gta{\mathrel{\spose{\lower 3pt\hbox{$\mathchar"218$}}
     \raise 2.0pt\hbox{$\mathchar"13E$}}}

\newcommand{\etal}{{et al.\ }}

\def\kms{\,{\rm km\,s^{-1}}}

\def\cm3{\,{\rm cm^{-3}}}

\shortauthors{Madau, Shen, \& Governato}
\submitted{}


\begin{document}

\title{Dark Matter Heating and Early Core Formation in Dwarf Galaxies}
\author{Piero Madau$^1$, Sijing Shen$^1$, \& Fabio Governato$^2$} 
\altaffiltext{1}{Department of Astronomy and Astrophysics, University of California, 1156 High Street, Santa Cruz, CA 95064, USA.}
\altaffiltext{2}{Astronomy Department, University of Washington, Seattle, WA 98195, USA.}

\begin{abstract}
We present more results from a fully cosmological $\Lambda$CDM simulation of a group of isolated dwarf galaxies that 
has been shown to reproduce the observed stellar mass and cold gas content, resolved star formation histories, and 
metallicities of dwarfs in the Local Volume. Here we investigate the energetics and timetable of the cusp-core transformation. 
As suggested by previous work, supernova-driven gas outflows remove dark matter (DM) cusps and create kpc-size cores in all systems having a stellar 
mass $M_*>10^6\,\msun$. The ``DM core mass removal efficiency"  -- dark mass ejected per unit stellar mass -- ranges today from a few to a dozen,
and increases with decreasing host mass. Because dwarfs form the bulk of 
their stars prior to redshift 1 and the amount of work required for DM heating and core formation scales approximately as $\Mvir^{5/3}$, the unbinding of the DM 
cusp starts early and the formation of cored profiles is not as energetically onerous as previously claimed. DM particles in the cusp typically migrate to 2-3 
core radii after absorbing a few percent of the energy released by supernovae. The present-day slopes of the inner dark matter mass profiles, 
$\Gamma\equiv d\log M/d\log R\simeq 2.5-3$, of the simulated ``Bashful" and ``Doc" dwarfs are similar to those measured in the luminous Fornax and Sculptor 
dwarf spheroidals. None of the simulated galaxies has a circular velocity profile exceeding $20\,\kms$ in the inner 1 kpc, implying that supernova feedback 
is key to solve the ``too-big-to-fail" problem for Milky Way subhalos.  
\end{abstract}

\section{Introduction}

Over the last two decades, observations of dwarf galaxies have challenged in a variety of ways our understanding of the mapping 
from dark matter halos to their baryonic components \citep[see, e.g.,][]{Pontzen14,Primack12}. One of the dwarf galaxy puzzles that has emerged both 
in the field and in the Galactic environment is basically a structural mismatch. Dark matter-only 
simulations predict steep (``cuspy") inner density profiles, but the observed rotation curves of dwarf galaxies show the sign of a near-constant density core 
\citep{Moore94,Flores94,deBlok02,deBlok08,Kuzio08,Walker11,Agnello12}. While this may indicate the need for more complex physics in the dark sector itself, 
emerging evidence suggests that a poor understanding of the baryonic processes involved in galaxy formation may be at the origin of this dwarf ``core-cusp problem". 
Rapid mass loss driven by supernovae (SNe) has long been argued to reduce the baryonic content of luminous dwarfs \citep{Dekel86,Mori02,Governato07} and 
flatten their central DM cusps \citep{Read05,Mashchenko06,Mashchenko08}. A new generation of hydrodynamic simulations, with sufficient resolution to model clustered 
star formation and the impact of SN-driven winds on the central mass distribution of dwarfs \citep{Governato10,Governato12,Teyssier13,DiCintio14}, together with analytic 
modeling \citep{Pontzen12}, have shown that gravitational potential fluctuations tied to efficient SN feedback can flatten the central cusps of halos in the most massive 
dwarfs. It remains unclear, however, whether this solution may work in the dwarf spheroidal galaxies of the Local Group, which are quite deficient in stars and subject 
to structure-changing ram pressure stripping and tidal heating by the host halo and disk \citep{Zolotov12,Brooks12,Garrison-Kimmel13,Arraki14}. Moreover, the core-cusp problem and 
the ``missing satellite" problem \citep{Moore99,Klypin99}) appear to place conflicting demands on the mass-to-light 
ratios of dwarfs, as core formation requires a relatively efficient conversion of gas into stars, while solving the abundance mismatch of CDM requires the opposite 
 \citep{Penarrubia12}.

\begin{figure*}
\centering
\includegraphics[width=0.85\textwidth]{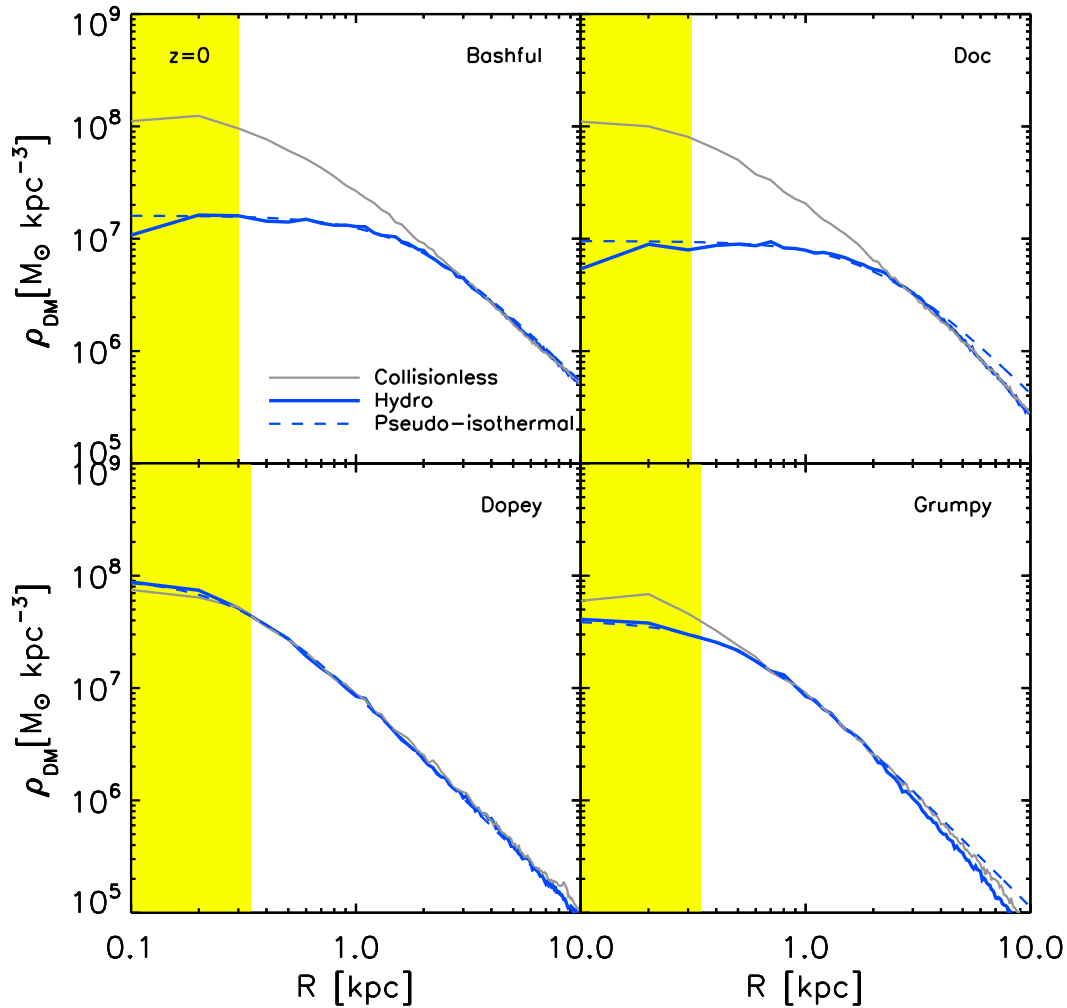}
\vspace{0.0cm}
\caption{The spherically-averaged (in 100 pc thick radial shells) density profile of the DM in four 
simulated dwarfs at $z=0$. {\it Solid blue curves:} fiducial hydrodynamic run with gas cooling, star 
formation, and gas outflows. {\it Solid gray curves:} control DM-only run. {\it Dashed blue curves:} 
pseudo-isothermal halo fits. The yellow shading indicates the region within which numerical convergence 
is not achieved because of two-body relaxation \citep{Power03}. No core is observed in Dopey, which forms 
a total of only $10^5\,\msun$ stars. 
}
\label{fig1}
\end{figure*}

To shed light on these controversies, we have recently run a fully cosmological $\Lambda$CDM simulation of a group of 
seven field dwarf galaxies with present-day virial masses in the range $\Mvir=4.4\times 10^8-3.6\times 10^{10}\,\msun$ \citep{Shen14}.
Our simulation -- one of the highest resolution ``zoom-ins" of field dwarfs run to redshift 0 -- was performed with the 
parallel TreeSPH code \textsc{Gasoline} \citep{Wadsley04} using approximately 6 million dark matter and an equal number of SPH particles. 
The gravitational spline softening length for collisional and collisionless particles was fixed to $\epsilon_G=86$ pc (physical). In high-density 
regions the gas smoothing length was allowed to shrink to $0.1\,\epsilon_G$ to ensure that hydrodynamic forces are resolved on tens of parsecs scales.
The simulation includes a star formation recipe based on a high gas density threshold, metal-dependent radiative cooling, a sub-grid scheme for the turbulent 
diffusion of thermal energy and metals, and a spatially uniform photoionizing UV background. 
To generate large-scale galactic outflows, we used the ``delayed radiative cooling" scheme of \citet{Stinson06}. The modeling of recurring SNe by the Sedov solution 
is best interpreted as a simplified algorithm to mimic the effect of energy deposited in the local ISM by multiple, clustered sources of mechanical luminosity. 
The importance of properly accounting for the entire momentum and energy budget of stellar feedback, including momentum injection by radiation pressure and stellar winds 
from massive stars, has been recently  stressed by many authors \citep[see, e.g.][and references therein]{Agertz13,Hopkins11,Hopkins13}. \citet{Agertz13} have shown, in particular, 
that simulations with maximal momentum injection suppress star formation to a similar degree as found in simulations that, like ours, adopt adiabatic thermal feedback. 
Note, incidentally, that our feedback model is not strictly speaking adiabatic, since SN-heated gas particles can exchange thermal energy with the ambient medium via
turbulent mixing \citep{Shen13}, and can therefore cool ``indirectly". A new superbubble-based feedback scheme for galaxy simulations \citep{Keller14} results in dwarfs 
with comparable mass-loaded outflows than the delayed radiative-cooling feedback method adopted here.    

As a validation of our approach, we have previously shown that our 
simulations are able to simultaneously reproduce the main current observables in low-mass systems, from their stellar mass and cold gas content, through their 
resolved star formation histories, to their gas-phase and stellar metallicities \citep{Shen14}. 
In this {\it Letter}, we focus instead on the effect of SN feedback on the inner density profiles of the simulated dwarfs and on the
energetics and timetable of the cusp-core transformation.

\begin{table*}[htp]
\centering
\caption{\enspace Present-Day Properties of Simulated Dwarfs}\label{tab:7dwarfs}
\begin{tabular*}{\hsize}{@{\extracolsep{\fill}}lccccccccc}
\\[-5pt]
\hline
\hline
\multicolumn{1}{l}{Name} &
\multicolumn{1}{c}{$\Mvir$} &
\multicolumn{1}{c}{$\Rvir$} &
\multicolumn{1}{c}{$V_{\rm max}$} &
\multicolumn{1}{c}{$M_*$} &
\multicolumn{1}{c}{$M_{\rm gas}$} &
\multicolumn{1}{c}{$\Delta E_{\rm SN}$} &
\multicolumn{1}{c}{$R_c$} &
\multicolumn{1}{c}{$\rho_0$} &
\multicolumn{1}{c}{$\Delta M_{\rm DM}/M_*$} \\
\multicolumn{1}{l}{} &
\multicolumn{1}{c}{[$\msun$]} &
\multicolumn{1}{c}{[kpc]} &
\multicolumn{1}{c}{[$\kms$]} &
\multicolumn{1}{c}{[$\msun$]} &
\multicolumn{1}{c}{[$\msun$]} &
\multicolumn{1}{c}{[ergs]} & 
\multicolumn{1}{c}{[kpc]} & 
\multicolumn{1}{c}{[$10^{-3}\msun$ pc$^{-3}$]} &
\multicolumn{1}{c}{} \\ 
\hline
\\[-5pt]
Bashful & $3.59\times 10^{10}$ & $85.23$ & $50.7$ & $1.15\times 10^8$ & $8.14\times 10^8$ & $2.1\times 10^{57}$ & $1.77$ & $17.5$ & 2.4\\
Doc     & $1.16\times 10^{10}$ & $50.52$ & $38.2$ & $3.40\times 10^7$ & $1.74\times 10^8$ & $6.1\times 10^{56}$ & $2.07$ & $10.2$ & 7.4\\
Dopey   & $3.30\times 10^{9}$ &  $38.45$ & $22.9$ & $9.60\times 10^4$ & $4.47\times 10^7$ & $1.6\times 10^{54}$ & \textemdash & \textemdash & \textemdash \\
Grumpy  & $1.78\times 10^{9}$ &  $29.36$ & $22.2$ & $5.30\times 10^5$ & $3.00\times 10^7$ & $8.5\times 10^{54}$ & $0.50$ & $48.5$ & 14.3\\
\hline
\end{tabular*}
\tablecomments{Column 1 lists the dwarf names. Columns 2, 3, 4, 5, 6, 7, 8, and 9 give the present-day virial mass, virial radius (defined as the radius 
enclosing a mean density of 93 times the critical density), maximum circular velocity, stellar mass, gas mass, 
total energy injected by core-collapse SNe, core size and central density of the pseudo-isothermal halo fit, and 
``DM core mass removal efficiency" (see text for details). Each Type II SN deposits an energy of $10^{51}\,$ergs into the gas, and the number of SNe per unit stellar
mass is $0.0115\,\msun^{-1}$. The DM and initial gas and star particle masses of the simulation are $m_{\rm DM}=1.6\times 10^4\,\msun$, 
$m_{\rm SPH}=3.3\times 10^3\,\msun$, and $m_*=10^3\,\msun$.
}
\vspace{0.5cm}
\end{table*}

\section{Present-Day Dark Matter Profiles}

Table \ref{tab:7dwarfs} summarizes the present-day properties of the four dwarfs that form stars in our simulation, named
``Bashful", ``Doc", ``Dopey", and ``Grumpy" in order of decreasing virial mass. 
They are characterized by blue colors, low star formation efficiencies, and high cold gas to stellar mass ratios \citep{Shen14}.
Their inner 2 kpc density profiles are resolved with as many as
$2.3\times 10^4$ DM particles (Bashful) and as few as $1.3\times 10^4$ (Grumpy), and their maximum circular velocities bracket the interval $22-51\,\kms$.
The two most massive dwarfs, Bashful and Doc, are within 100 kpc of each other and form an isolated  dwarf-dwarf galaxy pair like those found in the {Sloan Digital 
Sky Survey} by \citet{Geha12}.  All are ``field" dwarfs with the nearest massive halo ($M_{\rm vir}=2.5\times 10^{12}\,\msun$) more than 3 Mpc away.

\begin{figure*}
\centering
\includegraphics[width=0.85\textwidth]{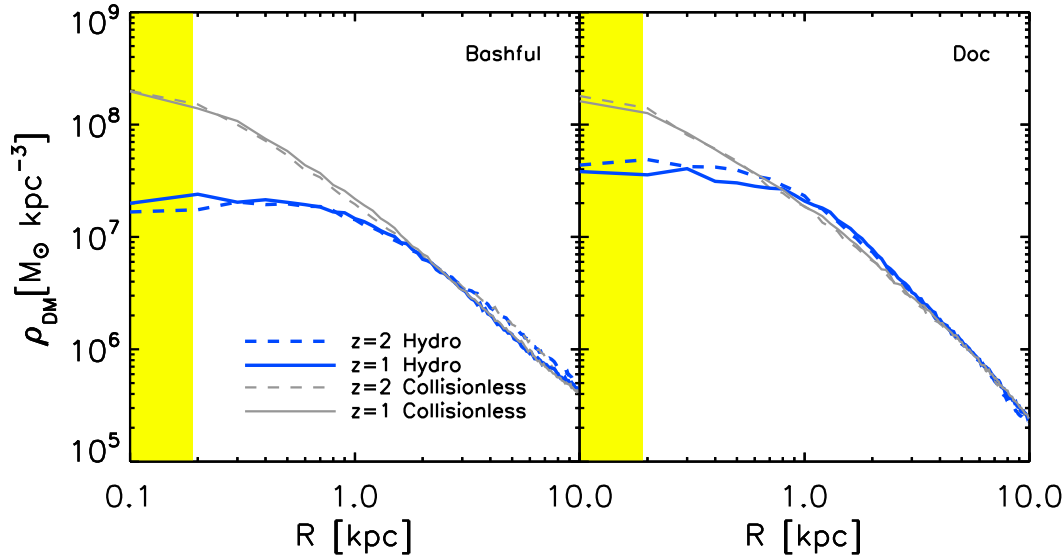}
\vspace{0.4cm}
\caption{Redshift evolution of the DM density profiles in the two most luminous simulated dwarfs.  {\it Blue curves:} profiles in the fiducial hydrodynamic
run at $z=1$ ({\it solid lines}) and $z=2$ ({\it dashed lines}). {\it Gray curves:} same in the control dissipantionless run.
The yellow shading indicates the region within which numerical convergence is not achieved at $z=1$ because of two-body relaxation.
}
\vspace{0.5cm}
\label{fig2}
\end{figure*}

We have fit the DM density distribution of each dwarfs with a ``pseudo-isothermal" profile defined as 
\begin{equation}
\rho_{\rm DM}={\rho_0\over 1+(R/R_c)^2}, 
\end{equation}
where $\rho_0$ is the central density and $R_c$ is the radius of the roughly constant density core. This empirically-motivated 
analytical function has proven to be a better description of both the shape of the rotation curve and the central densities of nearby 
dwarf galaxies than an NFW halo \citep{Kuzio08}, and to work well in the simulations of \citet{Teyssier13}. 
As can be seen from Figure \ref{fig1}, this is also true for our simulated dwarfs, for which a pseudo-isothermal 
fit is remarkably good within the central 5 kpc. At the present epoch, we measure core radii of 1.8 and 2.1 kpc in Doc and Bashful, 
respectively. A marginally-resolved core of $R_c=0.5$ kpc is also seen in Grumpy.
Note that mass models for nearby low surface brightness galaxies yield core radii that are often $\gta 1$ kpc \citep{Kuzio08}.  
The figure also shows the cuspy density profiles obtained in a {\it control DM-only simulation}. At the end of the hydrodynamic simulation, the 
dark matter density at the smallest resolved radius in Bashful (Doc) is 6 (14) times as small as the density obtained in the DM-only run. 
The offset between the inner (within 2 kpc) DM mass in the hydrodynamic (cored) and collisionless (cuspy) simulations, $\Delta M_{\rm DM}(R<2\,{\rm kpc})$, 
is $\Delta M_{\rm DM}=2.8\times 10^8\,\msun, 2.5\times 10^8\,\msun$ and $7.6\times 10^6\,\msun$ for Bashful, Doc, and Grumpy, respectively. {These mass deficits 
imply an ejected DM mass per stellar mass -- the {\it DM core mass removal efficiency} $\Delta M_{\rm DM}/M_*$ -- that ranges today from 2.4 in Bashful to 7.4 in Doc 
to 14.3 in Grumpy, i.e., that appears to increase with decreasing host mass.} 

The central densities of simulated halos can be artificially low due to two-body relaxation when the number of particles is low. 
Here we have conservatively applied the \citet{Power03} criterion to our DM-only runs to estimate the radius within which the two-body relaxation   
time is shorter than the Hubble time. We find convergence radii for our dwarfs between 300 and 350 pc 
at redshift 0, and close to 200 pc at redshift 1. No core is observed today in 
Dopey: in this very faint dwarf ($M_*\lta 10^5\,\msun$) the best-fit core radius is smaller than the numerical convergence radius and 
for this reason we do not report the parameters of the pseudo-isothermal halo fit in Table \ref{tab:7dwarfs}. 

\section{Cusp-Core Transformation}

Figure \ref{fig2} shows the DM density profiles in the two most luminous dwarfs, Bashful and Doc, at redshifts 1 and 2, as measured in the hydrodynamic run and in 
the control collisionless simulation. Grumpy was not included in the figure as it starts forming stars rather late, at $z<0.4$ \citep{Shen14}. Large cores are 
already in place at high redshift: the unbinding of the cusp starts early, when the host is significantly less massive than today (see also 
\citealt{Amorisco14,Gritschneder13,Mashchenko06}). We note that the formation of {\it early cores} must be a generic feature of dwarf galaxies in the real universe. 
The cumulative star formation histories of Bashful and Doc have been shown in \citet{Shen14} to be consistent with those inferred for dwarfs in the ACS Nearby 
Galaxy Survey Treasury program (ANGST) sample \citep{Weisz11} and based on color-magnitude diagrams of resolved stellar populations. The typical ANGST dwarf 
forms the bulk of its stars prior to redshift 1 and exhibits dominant ancient star formation ($>$ 10 Gyr ago) and lower levels of activity at intermediate 
times (1--10 Gyr ago). This simple fact underscores the danger of non-cosmological idealized calculations of the cusp-core transformation, where all the SN
energy is assumed to be released in the deeper gravitational potential of the dwarf descendant halo, rather than in the shallower well of an early progenitor. 

\begin{figure*}
\centering
\includegraphics[width=0.49\textwidth]{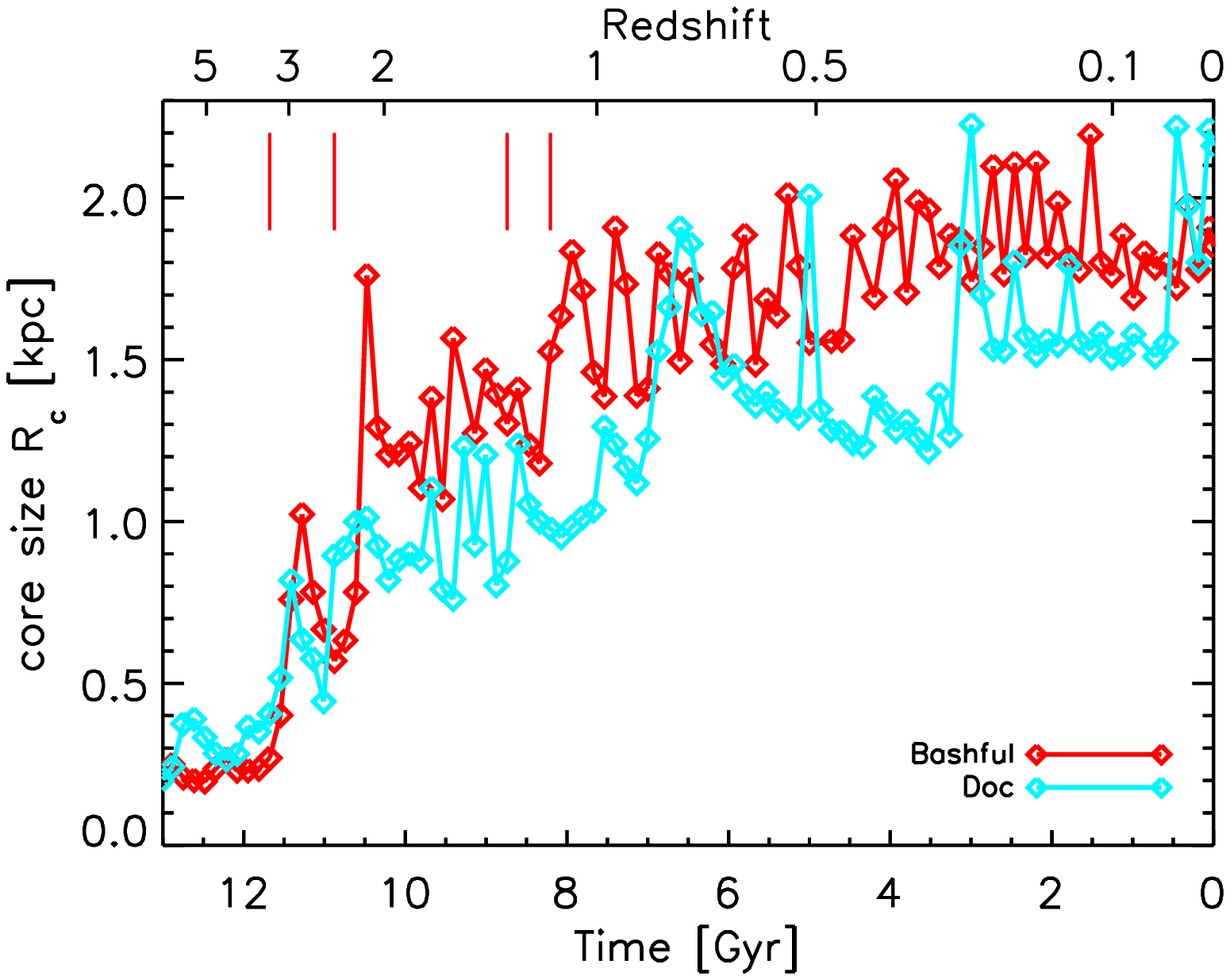}
\includegraphics[width=0.49\textwidth]{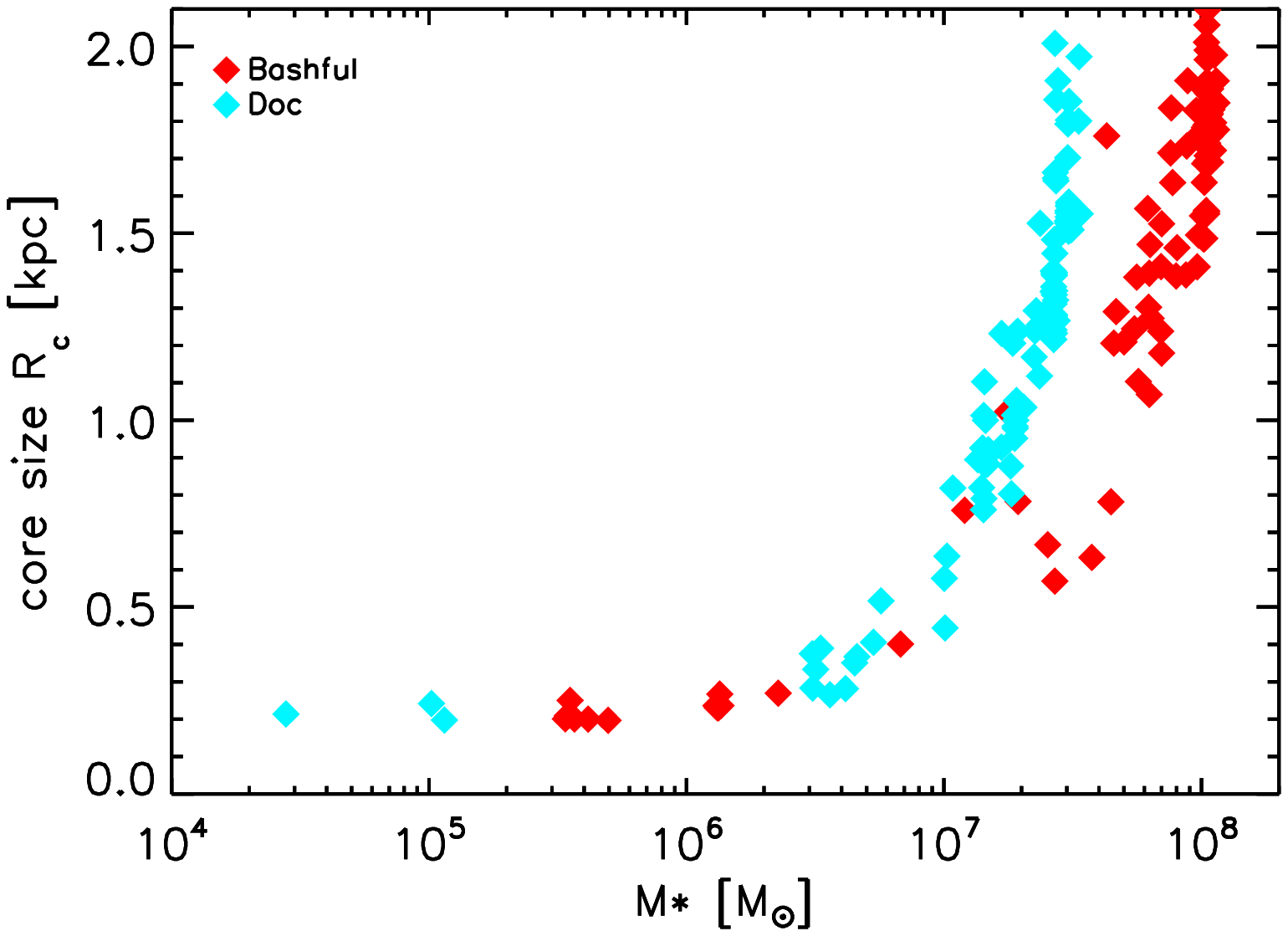}
\includegraphics[width=0.49\textwidth]{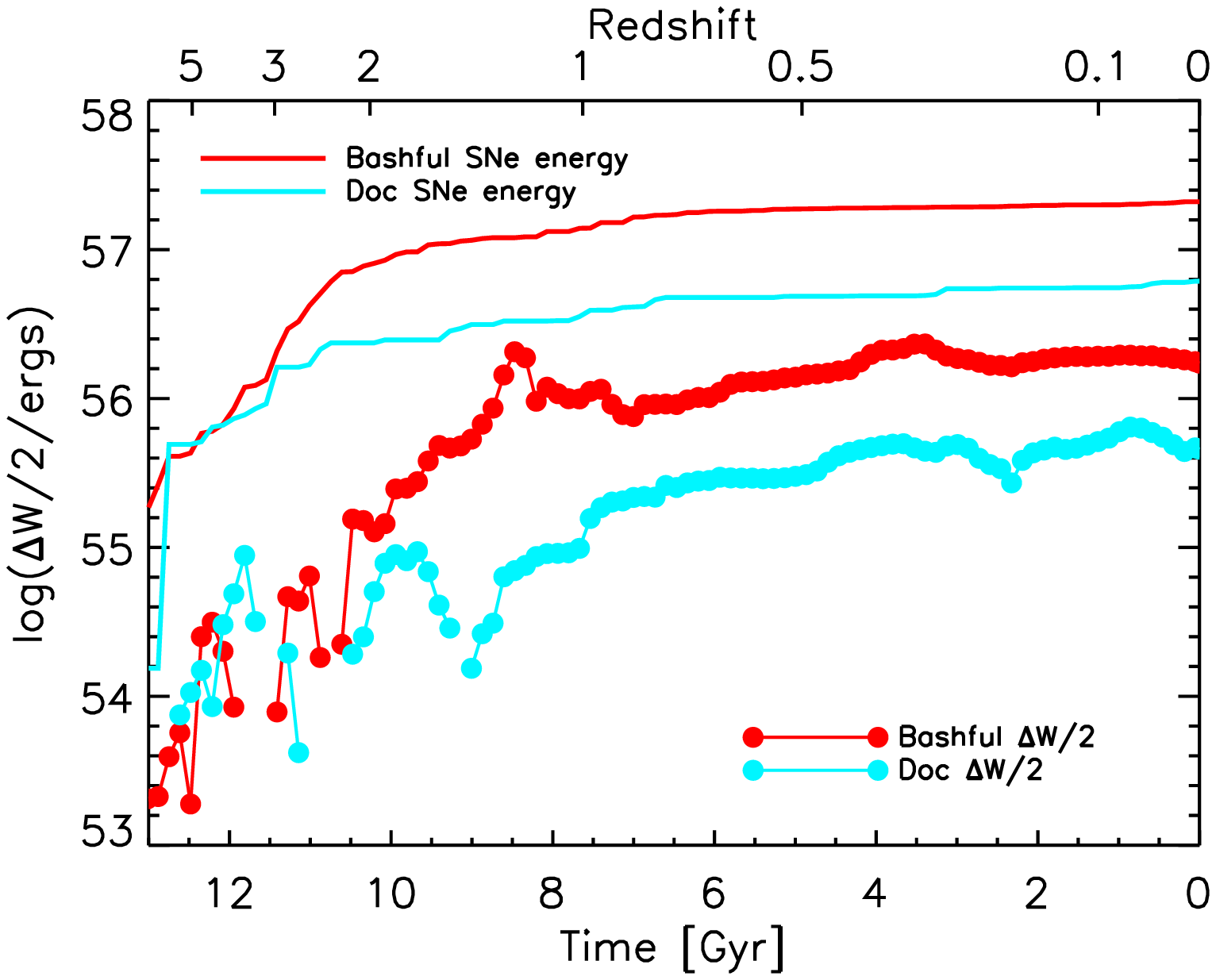}
\includegraphics[width=0.49\textwidth]{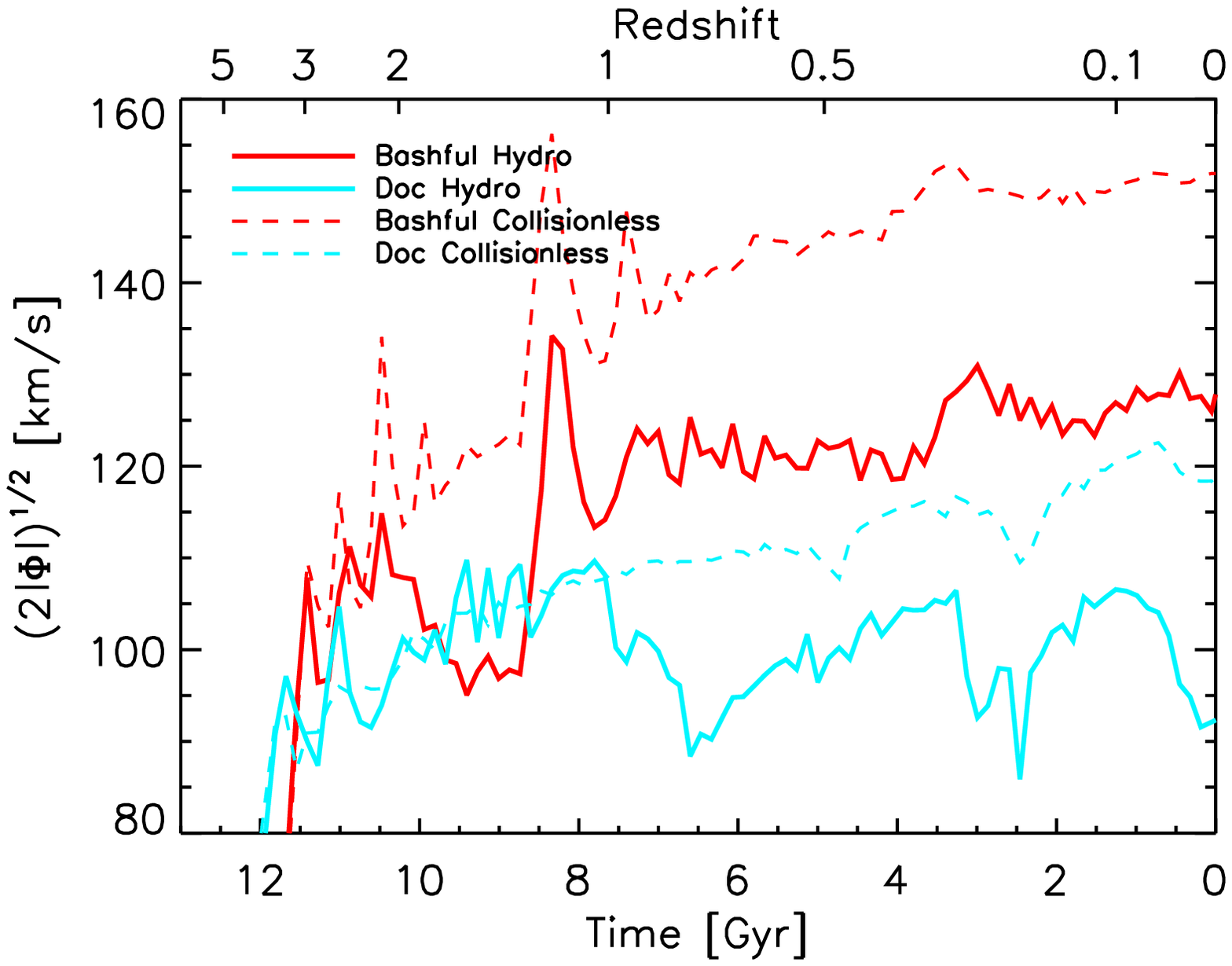}
\includegraphics[width=0.49\textwidth]{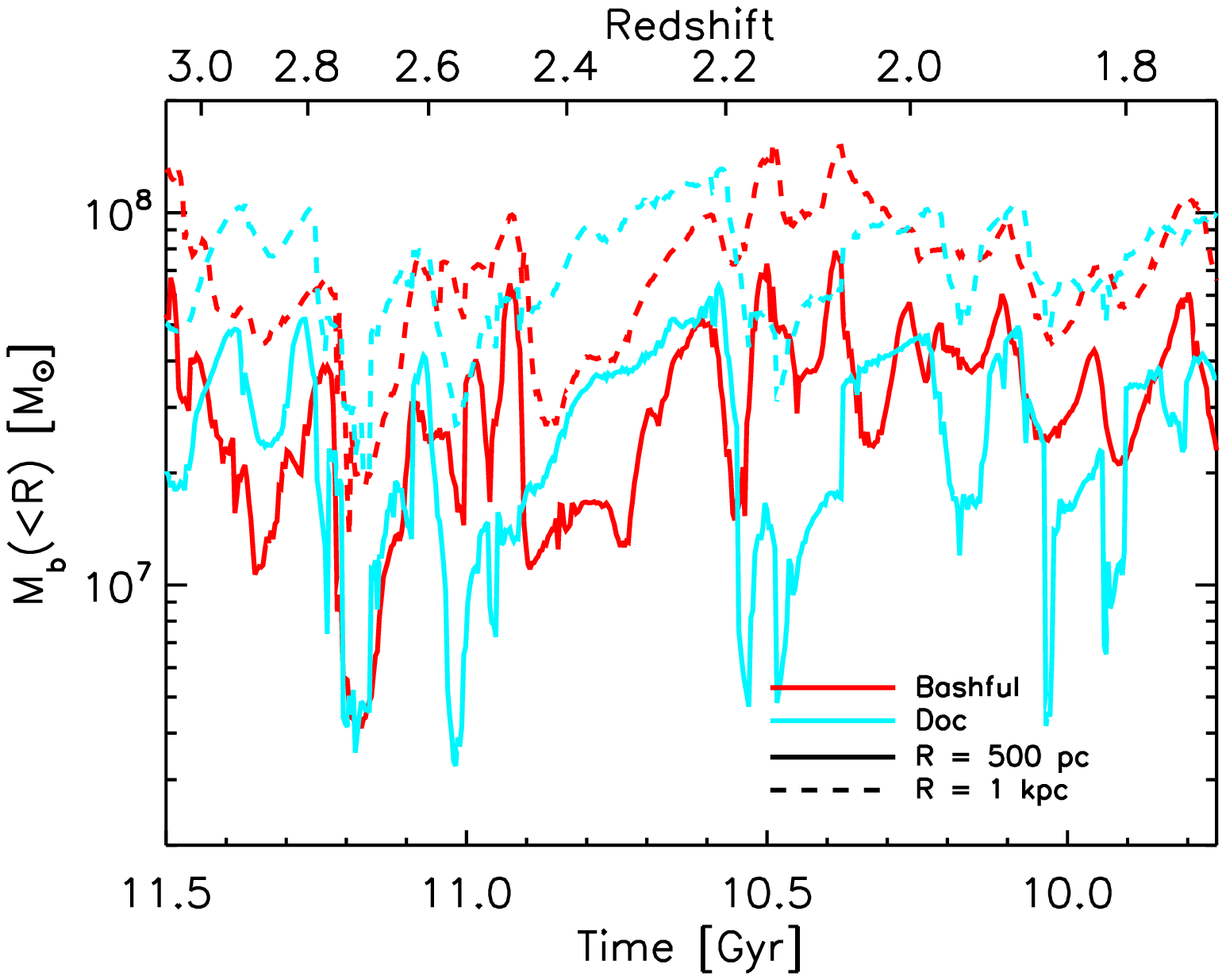}
\includegraphics[width=0.49\textwidth]{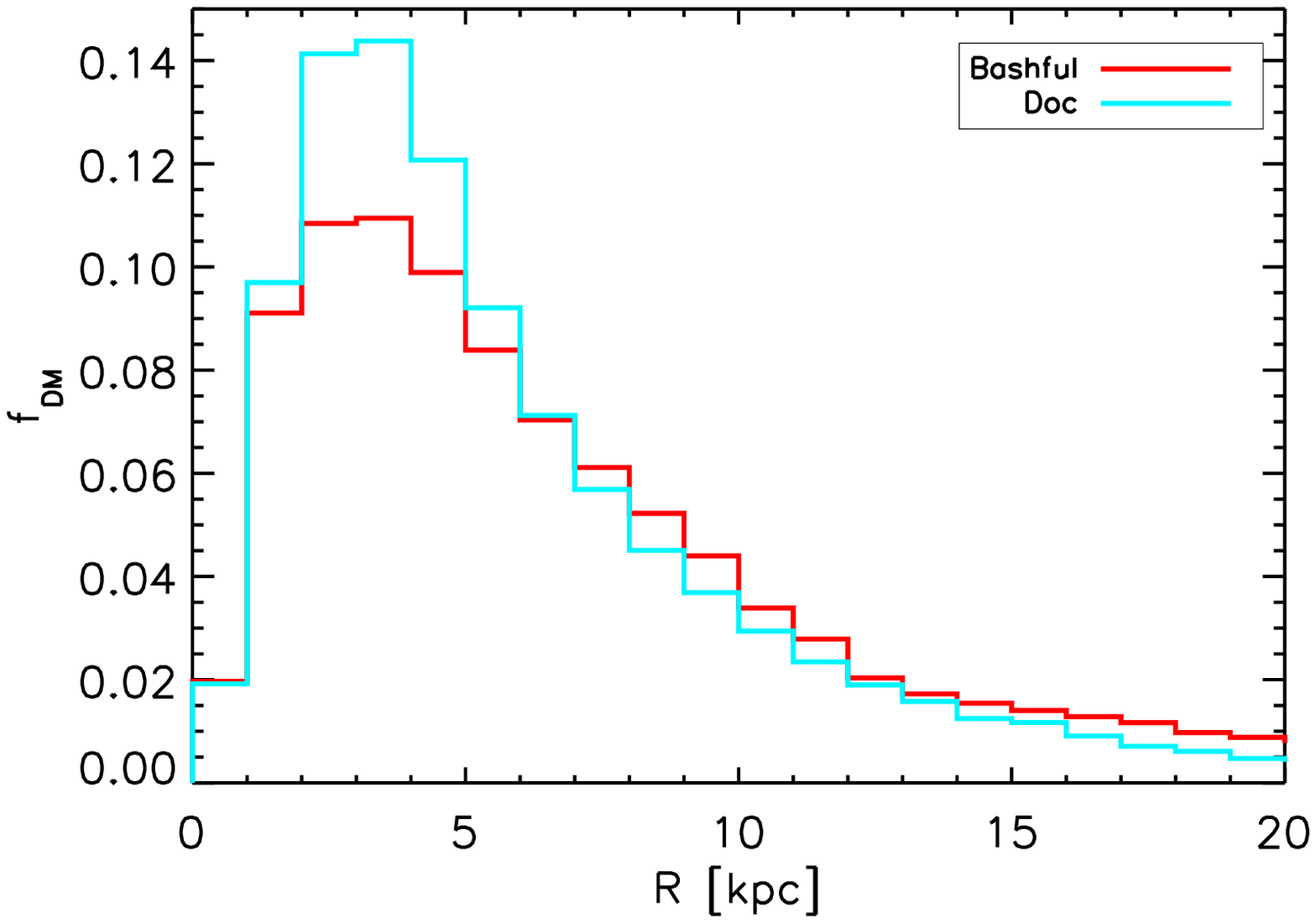}
\vspace{0.0cm}
\caption{Core-cusp transformation in Bashful and Doc. {\it Top left panel:} evolution of the DM core size $R_c$ with lookback time. The vertical marks show the location 
of the four most recent major ($>1:10$) mergers of Bashful. {\it Top right panel:} DM core size versus stellar mass. Core 
sizes greater than 1 kpc requires $M_*\gta 10^7\,\msun$, corresponding to a total energy injected by SNe of $\Delta E_{\rm SN}\gta 10^{56}$ ergs.
{\it Middle left panel:} minimum energy required for the cusp-core transformation, $\Delta W/2\equiv (W_{\rm core}-W_{\rm cusp})/2$, plotted against lookback time. 
Also shown ({\it solid lines}) in the total energy injected by SNe. 
{\it Middle right panel:} escape speed, $\sqrt {2|\phi(R=0)|}$, versus redshift from the center of Bashful and Doc in the hydrodynamical ({\it solid lines}) 
and DM-only ({\it dashed lines}) simulations. 
The late-time removal of material from a pre-existing DM core is not as energetically taxing as one would predict from DM-only simulations. 
{\it Bottom left panel:} baryonic mass interior to 500 pc (solid lines) and 1 kpc ({\it dashed lines}). Rapid fluctuations in the central potential are 
resolved on timescales $\gta$ 3.3 Myr. The enclosed baryonic mass shows the same cyclic behavior as the star formation rate in these systems.
{\it Bottom right panel:} normalized distribution of final ($z=0$) radii for all the DM particles that were once within 2 kpc from the center of Bashful and Doc. 
}
\label{fig3}
\end{figure*}

To shed more light on the energetics and timetable of the creation of cores in dwarf galaxies, we plot in the top panels of Figure \ref{fig3} the evolution with lookback time and 
stellar mass of the best-fit core radius $R_c$ in both Bashful and Doc. Cores on scales that are well resolved by our simulation form when $M_*>$ a few 
$\times 10^6\,\msun$. Core radii grow rapidly to kpc-size before redshift 2, and then oscillate around a 
value that increases slowly for 10 Gyr. The four vertical marks in the top left panel indicate the location of the most recent major ($>$1:10) mergers of 
Bashful: cores appear to be resilient and cusps are not {\it restored} following galaxy mergers. 

The left middle panel depicts the total energy, $\Delta E_{\rm SN}$, injected by Type II SNe as a 
function of time. \citet{Penarrubia12} used the virial theorem to estimate the minimum energy required for the cusp-core transformation as $\Delta W/2\equiv 
(W_{\rm core}-W_{\rm cusp})/2$, where $W$ is the gravitational binding energy of the DM halo,
\begin{equation}
W=-4\pi G\int_0^{\Rvir}\rho_{\rm DM}M(<R)RdR. 
\end{equation}    
Here, $M(<R)$ is the total enclosed mass, and we have computed $W_{\rm core}$ and $W_{\rm cusp}$ from the hydrodynamical and DM-only simulations, respectively. 
In the simulated dwarfs, $\Delta E_{\rm SN}\gg \Delta W/2$ at all times. At redshift 2, when the stellar masses of Bashful and Doc are $4.6\times 10^7\,\msun$ and $1.4\times 10^7\,\msun$, 
and their virial masses are 3.4 times and 2.2 times smaller than today, a pseudo-isothermal fit to their DM profiles yields core sizes of 1.3 kpc and 0.8 kpc, 
respectively. For reference, we measure at this epoch $(\Delta E_{\rm SN},\Delta W/2)=(8.1\times 10^{56}\,{\rm ergs}, 1.3\times 10^{55}\,{\rm ergs})$ in Bashful and 
$(2.4\times 10^{56}\,{\rm ergs}, 5.0\times 10^{54}\,{\rm ergs})$ in Doc. 
At later times, $z\lta 1.5$, while the host mass continues to grow, the ``scouring" of the 
{\it pre-existing} core by SN feedback progresses as DM particles are lifted from a potential that is significantly shallower 
than predicted by collisionless simulations (middle right panel in the figure). 

Our dwarfs have bursty star formation histories that are characterized by peak specific star formation rates in excess of 50--100 Gyr$^{-1}$, far outside the 
realm of normal, more massive galaxies. The bottom left panel of Figure \ref{fig3} shows the baryon mass interior to 0.5 and 1 kpc as a function of 
lookback time, with a fine-grained time resolution of 3.3 Myr. Impulsive mass-loss phases, punctuated by slower gas re-accretion, 
produce order-of-magnitude variations in the central potential on timescales that are shorter or comparable to the crossing time for DM particles 
(50--100 Myr at 0.5--1 kpc), creating an efficient channel for transferring kinetic energy from the gas to the DM, causing dark matter to migrate 
irreversibly outwards, and transforming a central DM density cusp into a near-constant density core \citep{Read05,Mashchenko08,Pontzen12}. 
Note that DM particles belonging to the central cusp do not become unbound but are typically displaced into 
the halo, with a distribution function that peaks today at $\sim$ 4--5 kpc from the center (bottom right panel in the figure). 

\section{Discussion}

Our simulations of realistic field dwarfs confirm that large DM cores are an unavoidable consequence of bursty star formation, 
shed light on the energetics and timetable of the cusp-core transformation, and show that SN feedback can effectively remove DM 
cusps in systems with $M_*> 10^6\,\msun$. Because luminous dwarfs form the bulk of their stars prior to redshift 1 and the 
amount of work required to remove a DM cusp scales approximately as $W\propto \Mvir^{5/3}$, the unbinding of the cusp starts 
early and the formation of cored DM profiles is energetically less demanding than estimated in non-cosmological
idealized calculations. We note that a stellar initial mass function that is shallower than the \citet{Kroupa01} functional form assumed in our simulations, 
like that recently measured in two ultra-faint dwarfs by \citet{Geha13}, would increase the SN feedback energy reservoir even further. 
Measurements of cored DM mass profiles in field dwarfs with $M_*< 10^6\,\msun$ appear then to be needed to provide a more powerful challenge to 
CDM-based models. 

It is also important to stress at this point that solutions to the core-cusp problem rest crucially upon the efficiency with which stars form in DM halos, and that
the $z=0$ stellar mass fractions, $M_*/\Mvir\simeq 0.003$, of Bashful and Doc are in excellent agreement with the stellar mass-halo mass (SMHM) relation for 
present-day dwarfs derived by \citet{Behroozi13}.  The low mass power-law behavior of the SMHM relation is broken on dwarf galaxy scales, corresponding to an 
upturn in the stellar mass function below $10^{8.5}\,\msun$ \citep{Baldry08}. Such an upturn at low stellar masses implies that dwarf galaxies have higher stellar mass 
fractions than predicted by assuming a scale-free power law below $\Mvir=10^{11}\,\msun$. {\it In particular, an $\Mvir\approx 10^{10}\,\msun$ halo at $z=0$ 
(like Doc) may be expected to form in excess of $10^7\,\msun$ in stars. This is an order of magnitude larger than assumed on these mass scales by \citet{Penarrubia12}.}

\begin{figure*}
\centering
\includegraphics[width=0.49\textwidth]{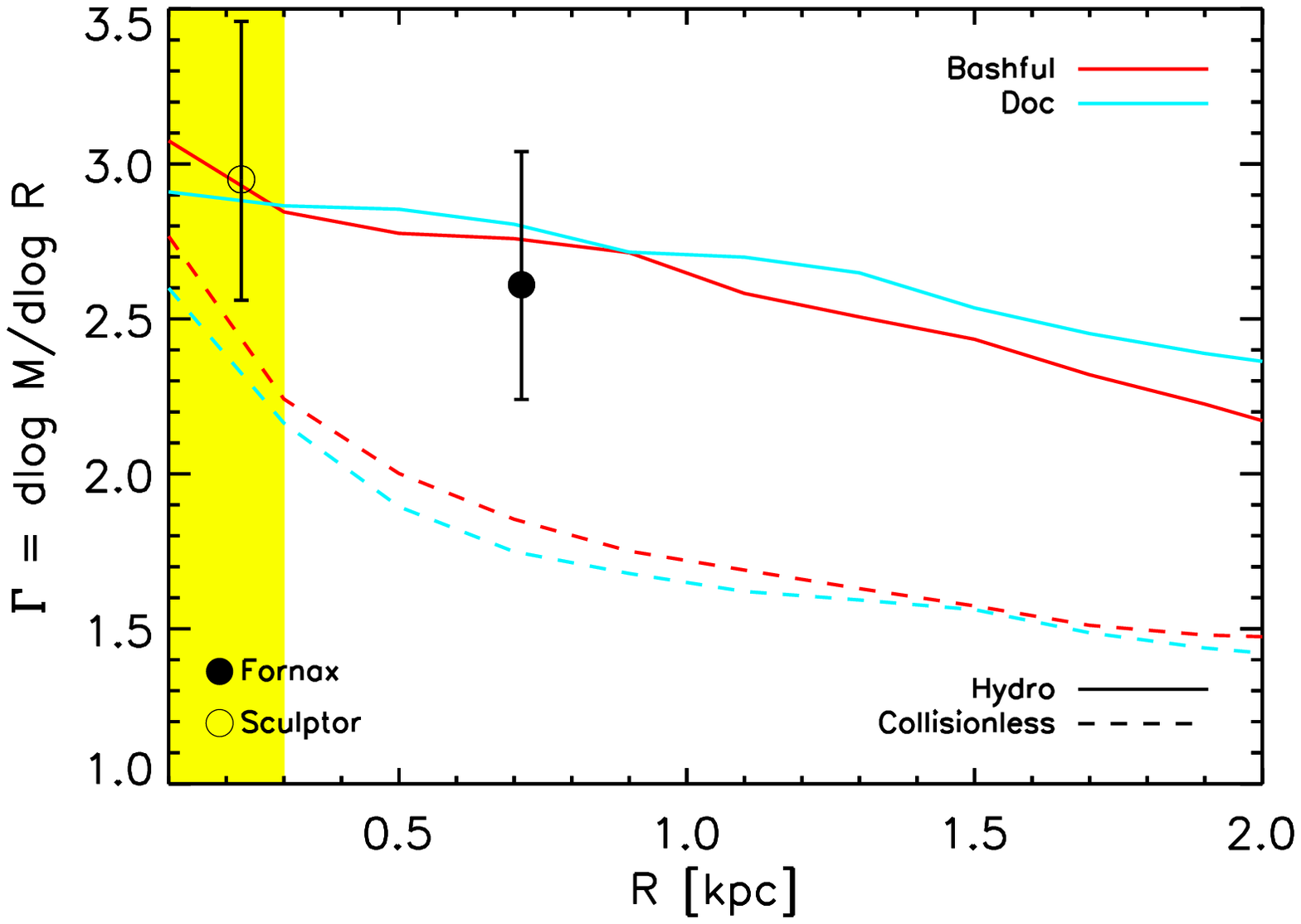}
\includegraphics[width=0.49\textwidth]{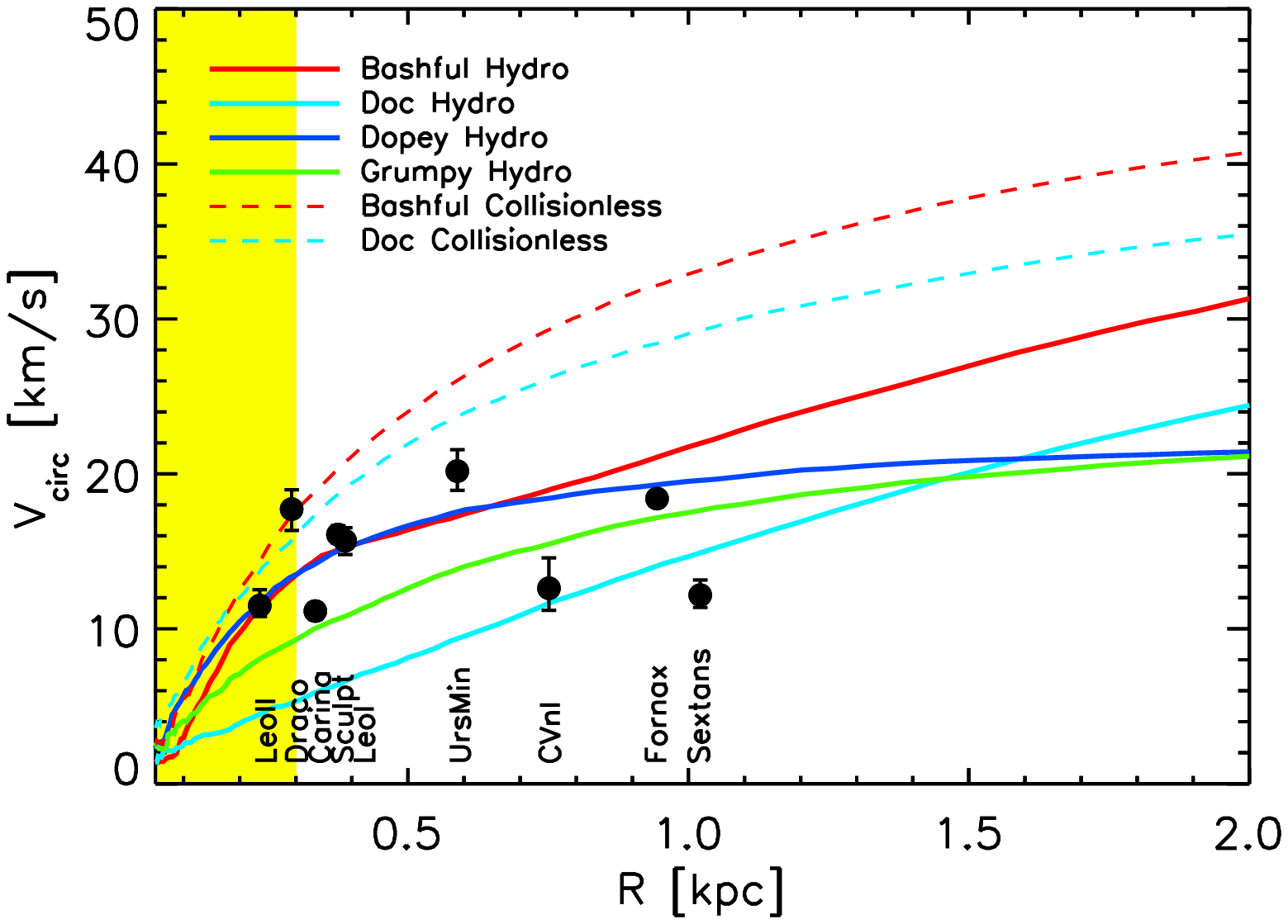}
\vspace{0.0cm}
\caption{Supernova feedback and the properties of dSphs. {\it Left panel}: slope of the mass profile $\Gamma\equiv d\log M/d\log R$
as a function of radius for the simulated Basshful and Doc dwarfs. {\it Solid lines}: hydrodynamical run. {\it Dashed lines}: control DM-only run.
NFW and cored halos have $\Gamma<2$ and $\Gamma<3$, respectively. Recent measurements of $\Gamma$ in the
Fornax and Sculptor dSphs by \citet{Walker11} are also shown.
{\it Right panel}: ``too-big-to-fail problem". The circular velocity profiles as a function of radius for the four simulated dwarfs Bashful, Doc, Dopey, and Grumpy
are shown with the solid lines. For Bashful and Doc we also show the velocity profiles obtained in the control DM-only run ({\it dashed lines}). 
None of the four simulated dwarfs has a circular velocity above $20\,\kms$ in the inner 1 kpc.
The points show the values at the half-light radius inferred in nine bright Milky Way dSphs by \citet{Wolf10}. Milky Way satellites do not have
significantly less mass near the center than our simulated massive dwarfs.
The yellow shading indicates the region within which numerical convergence is not achieved at $z=0$ in Bashful and Doc.
}
\label{fig4}
\end{figure*}

Bright dwarf spheroidal (dSph) galaxies in the halo of the Milky Way have estimated virial masses at infall well above 
$\Mvir=10^{9}\,\msun$ \citep[e.g.,][]{Rashkov12}. Our simulations may therefore offer a solution to some of the tension between $\Lambda$CDM 
predictions and detailed observations of these systems. \citet{Walker11} have recently introduced a method for measuring the slope of the total mass profile,
\begin{equation}
\Gamma\equiv d\log M/d\log R,
\end{equation} 
for dSphs that have chemodynamically distinct, equilibrium stellar subcomponents that independently trace the same 
spherical dark matter potential. Applied to published spectroscopic data for the Fornax and Sculptor dSphs, this technique yields slopes of 
$\Gamma=2.61^{+0.43}_{-0.37}$ and $\Gamma=2.95^{+0.51}_{-0.39}$, respectively. These values are consistent with cored dark matter halos of constant 
density over the central few hundred parsecs of each galaxy and rule out cuspy Navarro-Frenk-White (NFW) profiles ($d\log M/d\log R\le 2$ at all radii) 
with a high significance. With stellar masses of $M_*=(3.12\pm 0.35)\times 10^7\,\msun$ and $M_*=(8.0\pm 0.7)\times 10^6\,\msun$ \citep{Amorisco14}, 
Fornax and Sculptor are located in the region of the $R_c$--$M_*$ plane in Figure \ref{fig3} where 0.5--1 kpc cores are indeed expected to develop.
The left panel of Figure \ref{fig4} compares the $\Gamma$ values observed in these dSphs with those measured in the hydrodynamic and DM-only simulated 
profiles of Bashful and Doc, showing how SN feedback alone may explain the observed mass profiles of luminous dSphs.

Another version of the core-cusp puzzle in dSphs is the so-called ``too-big-to-fail" problem, which is based on the finding that the most 
massive subhalos in dissipationless simulations of Milky Way-size halos have too high central densities to host any of its bright satellites \citep{Boylan-Kolchin11}. 
In terms of subhalo maximum circular velocities and dSph circular velocity profiles, the problem can be simply stated as follows. A Milky Way-size halo is predicted 
to host several massive subhalos with maximum circular velocities today in excess of $30\,\kms$ \citep[e.g.,][]{Diemand08} --
the putative hosts of the brightest dSphs. Using subhalo profiles computed from DM-only simulations -- typically well-described by the NFW functional 
form -- one can then show that the circular velocities in the central regions of these subhalos exceed those inferred from stellar dynamics in the Milky Way dSphs 
\citep{Boylan-Kolchin12}. The right panel of Figure \ref{fig4} compares the circular velocity profiles of the four simulated dwarfs
with the circular velocities at the half-light radius inferred in nine bright Milky Way dSphs by \citet{Wolf10}, and  shows how SN feedback is key
to solve the ``too-big-to-fail" problem for Milky Way subhalos. Whereas in our control DM-only simulation, Bashful ($V_{\rm max}=51\,\kms$) and 
Doc ($V_{\rm max}=38\,\kms$) are indeed too dense to be compatible with any known satellite of the Milky Way, in the hydrodynamic run none of the simulated 
dwarfs has a circular velocity profile exceeding $20\,\kms$ in the inner 1 kpc, i.e., Milky Way satellites do not appear to have significantly less mass near the center 
than our simulated, $1.8\times 10^9\,\msun <M_{\rm vir}<3.6\times 10^{10}\,\msun$, field dwarfs.       

\acknowledgments
Support for this work was provided by the NSF through grants OIA--1124453 and AST--1229745, and by NASA through grant NNX12AF87G (P.M.). F.G. 
acknowledges support by the NSF through grants AST-0908499 and AST-1108885. We thank Joel Primack for a careful and critical reading of our manuscript.

\end{document}